\documentstyle[aps,epsf,rotate]{revtex}
\begin{document}

\bibliographystyle{prsty}
\draft
\title{Thermostating by deterministic scattering:\\ 
the periodic Lorentz gas}
\author{K. Rateitschak, R. Klages , G. Nicolis}
\address{Center for Nonlinear Phenomena and Complex Systems, 
Universit\'{e} Libre de Bruxelles, Campus Plaine CP 231, Blvd du
Triomphe, B-1050 Brussels, Belgium, krateits@ulb.ac.be}
\date{\today}
\maketitle
\renewcommand{\thefootnote}{\fnsymbol{footnote}}
\begin{abstract} 
We present a novel mechanism for thermalizing a system of particles in
equilibrium and nonequilibrium situations, based on specifically
modeling energy transfer at the boundaries via a microscopic collision process. We apply
our method to the periodic Lorentz gas, where a point particle moves
diffusively through an ensemble of hard disks arranged on a triangular
lattice. First, collision rules are defined for this system in thermal
equilibrium. They determine the velocity of the moving particle such
that the system is deterministic, time reversible, and
microcanonical. These collision rules can systematically be adapted to
the case where one associates arbitrarily many degrees of freedom to the disk, 
which here acts as a boundary. Subsequently, the system is
investigated in nonequilibrium situations by applying an external field. We show that in the limit where the disk is endowed by infinitely
many degrees of freedom it acts as a thermal reservoir yielding a
well--defined nonequilibrium steady state. The characteristic
properties of this state, as obtained from computer simulations, are
finally compared to the ones of the so--called Gaussian thermostated
driven Lorentz gas.
\end{abstract}

\pacs{PACS numbers: 05.20.-y, 05.45.+b, 05.60.+w, 05.70.Ln, 44.90.+c}

\section{Introduction}
The investigation of transport properties of many--particle systems in
nonequilibrium situations generally requires thermostats which remove excess 
energy to ensure the existence of nonequilibrium steady states
with constant, or on average constant, energy
\cite{HoAs75,HG83,AT87,EvMo90,Hoov91,Hess96,MoDe98}.
Hoover, Evans, Nos\'e and others developed methods of  
thermostating by
introducing a momentum--dependent friction coefficient into the
microscopic equations of motion, modeling the interaction of particles
with a thermal reservoir
\cite{EvMo90,Hoov91,HLM82,Ev83,EH83,Nose84a,Nose84b,Hoov85}. These
methods are deterministic and time reversible, in contrast to
stochastic thermostats \cite{AT87,LeSp78,TCG82,GKI85}. 
In this paper
we propose and analyze in detail an alternative deterministic
thermostat, based on including energy
transfer for a microscopic collision process between the moving particles and the boundaries instead of using a
momentum--dependent friction coefficient.
A short account of the main idea was reported in \cite{KRN98}. 

The two basic versions of a conventional deterministic thermostat are
the Gaussian thermostat and the Nos\'e--Hoover thermostat. The
Gaussian thermostat \cite{HLM82,Ev83,EH83} creates a microcanonical
ensemble in equilibrium and keeps the total energy (isoenergetic), or
the kinetic energy (isokinetic), constant in nonequilibrium. The
equations of motion of this thermostat can be derived from Gauss'
principle of least constraint \cite{MoDe98,EH83}.  The Nos\'e--Hoover
thermostat
\cite{Nose84a,Nose84b,Hoov85} creates a canonical ensemble in thermal
equilibrium and keeps the energy on average constant in
nonequilibrium.

Though the microscopic equations of deterministic thermostated systems
are time reversible the macroscopic dynamics is irreversible in
nonequilibrium leading to momentum and energy fluxes with well--defined 
transport coefficients \cite{HLM82,Ev83,EvHo85,MH87,HHP87}. Macroscopic
irreversibility is based on the fact that only one direction in time
is dynamically stable in these systems, whereas the time reversed
direction is dynamically unstable \cite{HHP87,Hoov88}.  This implies a
contraction of the phase space onto a fractal attractor during the forward 
evolution \cite{MH87,Morr87,Morr89,HoMo89,Mo89a,HooverPosch98,HooverPosch98a}.
In contrast, purely stochastic thermostats are expected to typically lead to 
a smooth phase space density in nonequilibrium \cite{GKI85,EyLe92}.  In
agreement with the phase space contraction, it has been found that the sum of 
the Lyapunov exponents is negative in deterministic
thermostated systems. It has been argued  by many authors that the rate of 
this phase space contraction is related to the thermodynamic entropy production
\cite{HHP87,PoHo87,PoHo88,PH89,Ch1,Ch2,ChLe95,TeVB96,VTB97,ChLe97,GaCo95a,Ruelle96}.
Furthermore, relations between the sum of the Lyapunov exponents and
the corresponding transport coefficient have been derived 
\cite{MH87,PoHo88,Ch1,Ch2,ECM,Vanc,BarEC,DeGP95}. In short,
deterministic thermostats provide an important approach to modeling 
nonequilibrium steady states and establishing interesting
links between dynamical system theory and statistical mechanics
\cite{EvMo90,Hoov91,Do99,TGN98,MaHo92,Mare97}.

On the other hand, conventional thermostats are based on a drastic modification of
the microscopic equations of motion by including momentum--dependent friction
coefficients, which implies that the microscopic equations cannot be
Hamiltonian anymore in their usual physical coordinates. Although 
there exist methods to relate them to generalized Hamiltonian systems by 
noncanonical transformations
\cite{MoDe98,Ho88,DeMo96,DettmannMorriss97,Choq98} the question
still remains whether the results obtained from these deterministic
thermostats provide general characteristics of nonequilibrium
steady states, or whether they depend on this particular way of
thermostating \cite{TGN98,MaHo92,Mare97}.

As an alternative, a specific  mechanism to simulate a steady shear flow 
without using a thermostat of the above kind has been studied in 
Refs.~\cite{ChLe95,ChLe97,DePo97b}. Here, the collision of
a particle with a wall is described by rules which change the scattering angle
but not the absolute value of the velocity of the particle.  Open systems 
with fixed concentration gradients at the boundaries have been the subject of 
another approach to 
create a nonequilibrium steady state \cite{GN,GaDo95,DoGa95,Gasp}. A
possible link of this approach to thermostated systems is discussed in Refs.~\cite{TeVB96,VTB97}.

A simple deterministic one--particle system in which there is evident need of
thermostating  is the field driven periodic Lorentz gas. 
We recall that the original Lorentz gas model 
consists of a system of randomly distributed hard disks and a particle that moves freely between successive elastic collisions with the disks \cite{Lo05}. Later, a periodic configuration of disks onto a triangular lattice known as the periodic Lorentz gas
\cite{BuSi81} has been introduced, and still serves as
a standard model in the field of chaos and transport
\cite{GN,Gasp,BuSi81,CvGS92,MaZw83,Gas93,GaBa94,MoRo94,Gasp96,MaMa97}.
In case of the driven Lorentz gas, an external electric field drives
the system into nonequlibrium by accelerating the moving
particle while pumping at the same time energy into the system through Joule heating.
A number of authors developed mechanisms for removing energy from this system through a Gaussian
isokinetic thermostat, which creates a nonequilibrium steady state
with constant energy of the particle
\cite{MH87,Ch1,Ch2,Vanc,BarEC,DeGP95,LRM94,LNRM95,DettmannMorriss96,DeMo97}. 
This thermostated one--particle system shows the same characteristics in
nonequilibrium as other many--particle systems: the phase space
density contracts onto a fractal attractor
\cite{MH87,LRM94,LNRM95,DettmannMorriss96}, and the sum of Lyapunov exponents 
is negative. Relations were derived between this sum of Lyapunov exponents, the conductivity and the irreversible entropy production of this system.
\cite{Ch1,Ch2,Vanc,BarEC,DeGP95,LNRM95,DeMo97}.
 We note that a model almost identical to the driven periodic Lorentz gas, except for some geometric restrictions, is the Galton board, which has been invented in 1873 to study probability distributions \cite{HooverPosch98}.

In the present paper we introduce an alternative method of
deterministic thermostating which is free of addition of new terms in the equations of motion, and illustrate it on the periodic Lorentz gas. 
The paper is organized as follows. In Section II we introduce our model in 
equilibrium. We define collision rules for the particle
which change its velocity at a collision with a scatterer such that the 
dynamics of the system is deterministic, time reversible, and yields the
microcanonical density in equilibrium. First the disk is equipped with
one degree of freedom, and energy conservation between particle and
disk determines the amount of energy on the disk after a collision.
Then the model is modified by pretending that the disk has arbitrarily
many degrees of freedom. This involves a redefinition of the
microscopic scattering rules to keep the dynamics microcanonical. 
The reduced densities
of an arbitrarily dimensional microcanonical system are then calculated. 
In Section III we numerically investigate the system in nonequilibrium by
switching on the external field. We show that in the limit of
associating infinitely many degrees of freedom to the disk our
mechanism keeps the energy of the moving particle constant on average,
thus leading to a well-defined nonequilibrium steady state. This
confirms that our mechanism yields indeed a proper thermostating. The
characteristic features of the resulting nonequilibrium steady state
for our model are discussed explicitly, especially in comparison to
the Gaussian thermostated periodic Lorentz gas. A summary with main
conclusions is given in Section IV.

\section{The model and its equilibrium properties} 
\begin{figure}
\epsfxsize=12cm
\centerline{\epsfbox{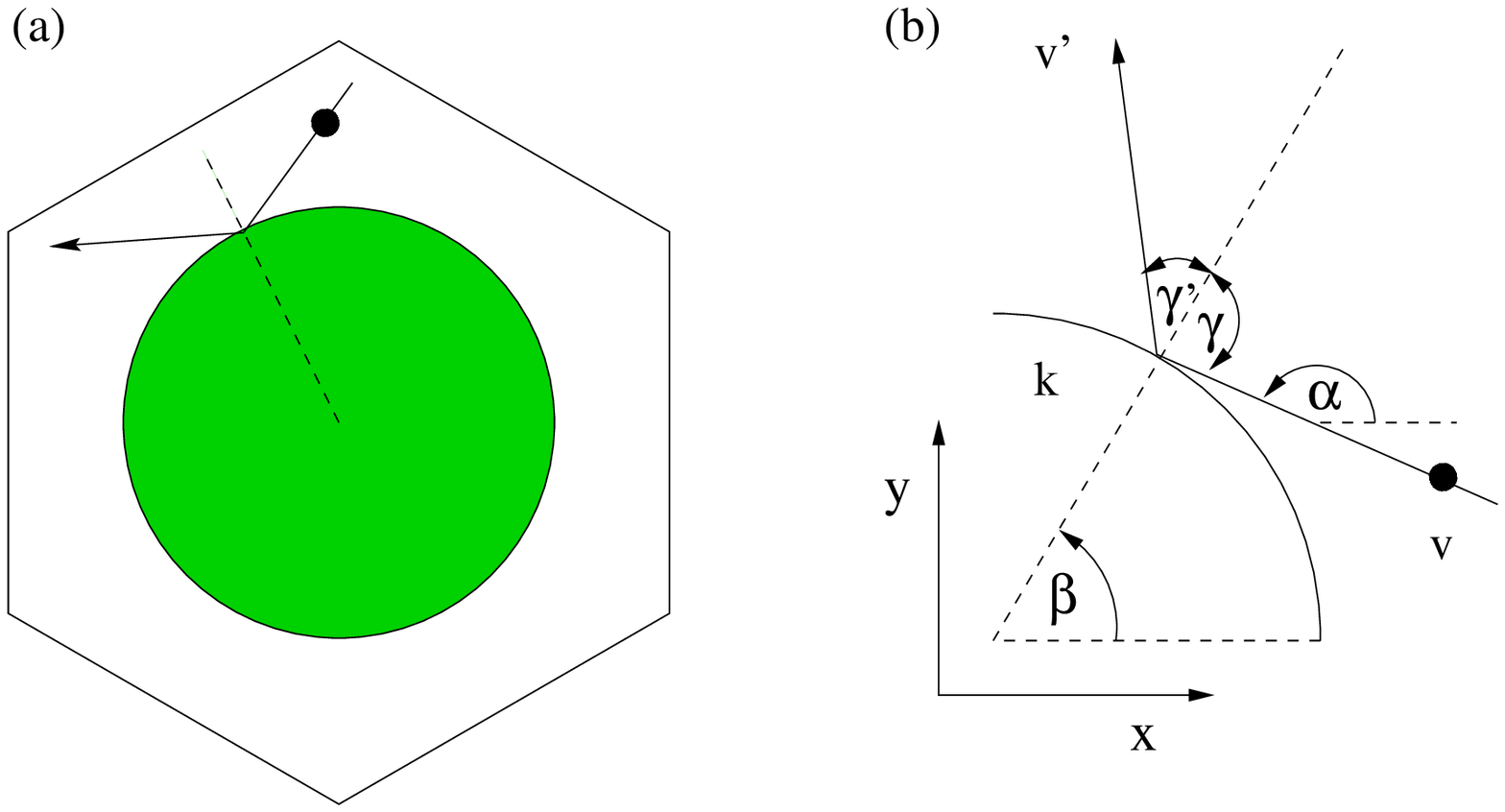}}
\vspace*{0.3cm} 
\caption{(a) Elementary cell of the periodic Lorentz gas on a
triangular lattice. (b) Definition of the relevant variables to
describe the collision process.}
\label{cell}
\end{figure}

Owing to the periodicity of the lattice, it will be sufficient to study the dynamics in one Lorentz gas cell with
periodic boundary conditions, see Fig.~\ref{cell}(a). As the radius of the disk we take $r=1$. For the
spacing between two neighboring disks we choose $w\simeq 0.2361$, as
is standard in the literature to ensure that a diffusion coefficient
exists \cite{MH87,MaZw83}. The variables intervening in the dynamics are defined in Fig.~\ref{cell}(b): $\beta$ is
the angular coordinate of the point at which the particle collides with the disk, $\gamma$ is the
angle of incidence at this point, and $\alpha$ is the angle of flight
of the particle.  The particle has two velocity components,
$v=(v_x,v_y)$.  To establish the energy transfer between particle and
disk we also assign to the disk a velocity $k$.  Thus, in total
our model has three degrees of freedom. To proceed further we now need to introduce 
specific
collision rules for the moving particle, which map $v$ and $\gamma$ onto 
$v'$ and $\gamma'$.  Energy conservation between particle and disk
then yields the velocity $k'$ onto the disk after a collision.

\subsection{Rotating disk model}
We first present a model based on a very simple physical mechanism
for a possible energy transfer between particle and disk at a
collision. This model has the required properties only in a limited
parameter range, but it contains some basic ideas for the formulation
of our thermostating mechanism, which we will introduce in
full detail afterwards.

The position of the disk is held fixed. The velocity of the
moving particle at a collision can be split into a normal component
$v_n$ and into a tangent component $v_t$. We now interpret the
velocity $k$ of the disk as a rotational degree of freedom.
This allows us to define a transfer of kinetic energy between $v_t$
and $k$ based straightforwardly on energy and momentum conservation,
while $v_n$ is elastically reflected. The collision rules thus
read
\begin{eqnarray}
v_t'&=&\frac{(m-m_k)v_t+2m_kk}{m+m_k}\nonumber\\
k'&=&\frac{(m_k-m)k+2mv_t}{m+m_k}\nonumber\\
v_n'&=&-v_n \nonumber\; ,
\end{eqnarray}
where $m$ is the mass of the particle and $m_k$ the mass of the disk.
 Computer simulations show that this model yields a
microcanonical probability density for a total kinetic energy of $E=0.5$, $m=m_k=1$ and large spacings between two neighboring disks $w\geq2.0$. However, by decreasing $w$ the
probability density starts to deviate from the microcanonical density
because the particle gets more and more trapped in parts of the
Lorentz gas cell. By increasing the mass of the disk $m_k$ deviations from the microcanonical density appear already for larger $w$. 

Thus, only for a certain choice of parameters is this simple approach leading to the desired result, which is that the system is microcanonical and shows equipartitioning of energy in all degrees of freedom, while, in general, the dynamics is apparently more complicated. In the
following we want to investigate whether by introducing more generic collision
rules we can achieve that the dynamics is microcanonical for any choice of respective parameters. Specifically the fact that the energy and momentum conservation law
is a linear two--dimensional map of the form $(v_t',k')=f(v_t,k)$ motivates us to define the collision process by a simple, chaotic two--dimensional map as discussed amply in the next section.
More details of the rotating disk model will be reported elsewhere \cite{RKrd}.

\subsection{Modeling the collision process by a two dimensional map}
\subsubsection{The baker map}
\label{zb}
We choose the well--known baker map
\cite{Do99}, which we apply to the variables
$(x_b,y_b)=(\sin|\gamma|,v)$. Here $x_b=\sin(|\gamma|)$ is the Birkhoff coordinate\footnote{Note that the Poincar\'e or Birkhoff mapping between two collisions is just area preserving in the Birkhoff coordinates $(\beta,\sin(\gamma))$.}
of $\gamma$ leading to $\varrho(\sin(|\gamma)|)\equiv 1$ \cite{Gasp}.
The change of these variables at a collision to
$({x_b}',{y_b}')=(\sin |\gamma'|,v')$ (see Fig.~\ref{cell}) is thus given by
\begin{equation}
(x_b',y_b') = M(x_b,y_b)  = \left\{ 
\begin{array}{r@{,}l@{\:,\:}l}
(2x_b & y_b/2) & x_b\le 0.5 \\ 
(2x_b -1& (y_b+1)/2) & x_b> 0.5 \\ 
\end{array} 
\right. \quad ,
\label{bk} 
\end{equation}
where $0\le y_b<1$. 
$k'$ is then obtained from energy conservation. Since $k$ is not
explicitly contained anymore in the collision rules given by Eq.~(\ref{bk}), 
one can argue that the detailed dynamics of $k$  is no longer
relevant for the moving particle. In particular, 
$k$ need no more be associated to a rotational degree of
freedom. Based on our general formulation of the collision rules its
physical interpretation as a degree of freedom is now more flexible. 
For example, one may think of $k$ as being related to some kind of lattice modes.

To ensure that the system is time-reversible, we let the forward baker act
if $0\le\gamma\le \pi/2$, and its inverse if $-\pi/2\le\gamma<0$. The
angle $\gamma'$ always goes to the respective other side of the
normal, that means $\gamma'$ has the opposite sign of $\gamma$, as shown 
in Fig.\ref{cell}. To avoid a symmetry breaking in a
possible nonequilibrium situation we alternate the assignment of the
forward and backward baker, that is, if  
$\mbox{int}(\beta\cdot 10^8)$ is even (odd) we take the forward (backward) 
baker for $\gamma\ge0$ and {\em vice versa} the corresponding backward 
(forward) baker for $\gamma<0$. 
Ideally, the alternation should be done in infinitely fine steps, which is not feasible in computer simulations.

\subsubsection{Relation between map density and time continuous density}
By investigating the dynamics of the collision process through a
baker map, one is actually considering the Poincar\'e section of the velocity 
of the particle at the moment of the collision. 
We denote the corresponding probability
density of the moving particle as the map density $\varrho_{map}(v)$.
$\varrho_{map}(v)$ can be written in discretized form as
$\varrho_{map}(v_i)=(\ldots,c_{v_i}/c_o,\ldots)$, where
$c_{v_i}$ is the number of collisions after which the particle has the
velocity $v_i$ and $c_o$ is the total number of collisions.

One may establish a relation between $\varrho_{map}(v)$
and the time continuous probability density $\varrho(v)$, where $v$ is
measured at any time interval dt. This is the relevant quantity to
check for a microcanonical distribution. 
Notice that we can write in the same way
as before $\varrho(v_i)=(\ldots,t_{v_i}/t,\ldots)$, where
$t_{v_i}$ is the total time during which the particle has the
velocity $v_i$, and t is the total time.

We introduce now the  mean time of flight between two collisions $<t>$ by
\begin{equation}
<t> = \frac{t}{c_o}
\label{at}
\end{equation}
and the mean time of flight between two collisions $<t>_{v_i}$ when
the particle has the velocity $v_i$ by
\begin{equation}
<t>_{v_i}= \frac{t_{v_i}}{c_{v_i}}=\frac{<s>}{v_i}\;.
\label{atv}
\end{equation}
$<s>$ is the collision length, which is expected to be independent of
$v$. To compute a value of $\varrho(v_i)$, Eq.~(\ref{at}) and
Eq.~(\ref{atv}) can be combined to
\begin{equation}
\frac{t_{v_i}}{t}=\frac{<s>}{<t>v_i}\frac{c_{v_i}}{c_0}\;.
\label{tv}
\end{equation}
Eq.~(\ref{tv}) is valid for all $i$ implying that we get the following
relation between the time continuous density $\varrho(v)$ and the map
density $\varrho_{map}(v)$,
\begin{equation}
\varrho(v) = \frac{\varrho_{map}(v)}{v}\frac{<s>}{<t>} = const.\frac{\varrho_{map}(v)}{v}\;,
\label{dt}
\end{equation}
where the constant is determined by normalization.
By assuming that the coupling of the collision rules of Eq.~(\ref{bk})
to the specific geometry of the Lorentz gas does not yield an
invariant map density being different from the one of the baker
map, we can now calculate $\varrho(v)$ for our model: Inserting $\varrho_{baker}\equiv 1$ in Eq.~(\ref{dt}) results in
$\varrho(v)=const./v$. This expression is not normalizable and
does not correspond to the correct result related to a microcanonical density. The way out of this difficulty is presented in the next section.

\subsubsection{Getting a microcanonical density}
\begin{figure}[htbp]
\epsfxsize=14cm
\centerline{\epsfbox{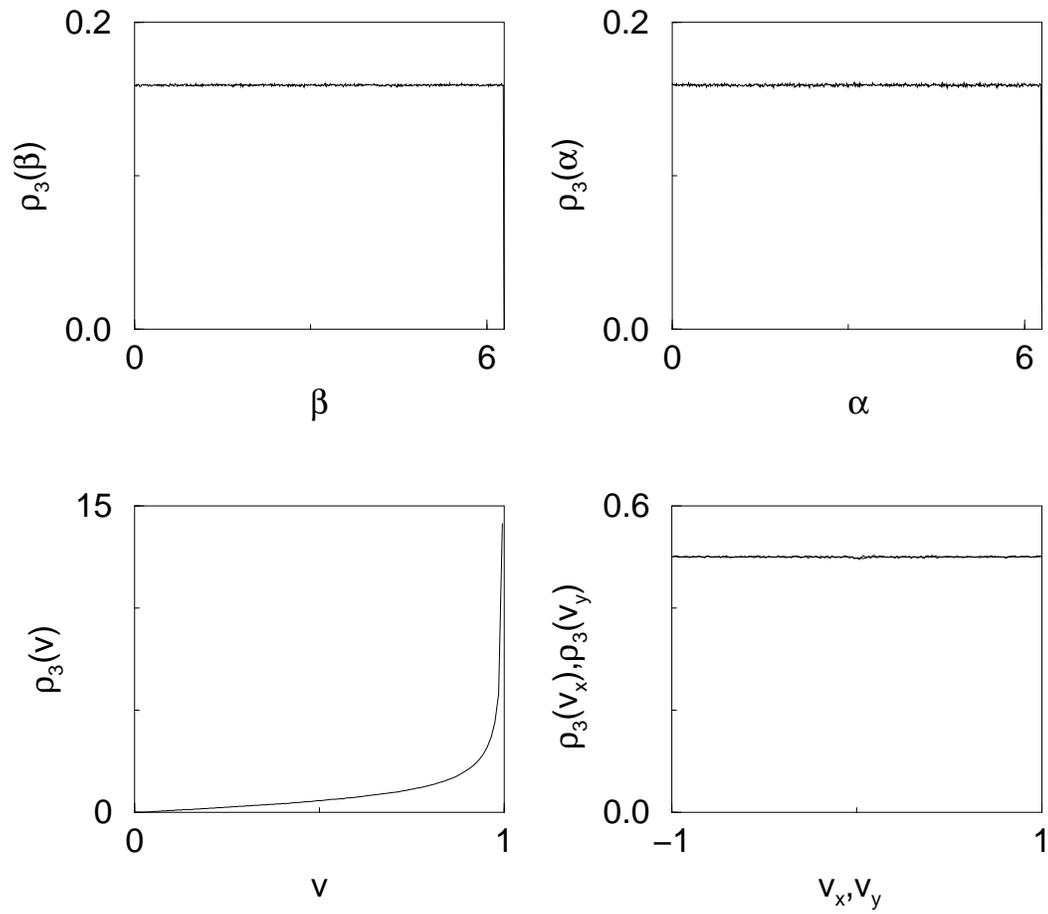}}
\caption{Probability densities for $d=3$ at $E=0.5$}
\label{figd3}
\end{figure}

Having shown in the last subsection that the collision rules as
described by the baker map are not sufficient to get a microcanonical
probability density, we now amend the definition
of the collision rules. We do this by including an additional
transformation $Y$ which is constructed in the following way: Our
system has three degrees of freedom $(v_x,v_y,k)$, and the total
energy $E$ is conserved, $2E={v_x}^2+{v_y}^2+k^2$. The dynamics is
microcanonical if the probability density is equidistributed on a
three--dimensional sphere,
\begin{equation}
\varrho_3(v_x,v_y,k)=\frac{1}{8\pi E}\delta(2E-{v_x}^2-{v_y}^2-k^2)\; .
\end{equation}
In the Appendix we calculate the reduced density for one or two degrees
of freedom of a $d$-dimensional microcanonical system.  For $d=3$,
Eq.~(\ref{dvx}) and Eq.~(\ref{dv}) lead to
\begin{eqnarray}
\varrho_3(v_x) &=& \frac{1}{2\sqrt{2E}} 
\label{dvx3}\\
\varrho_3(v) &=& \frac{v}{\sqrt{2E(2E-v^2)}}\;.
\label{dv3}
\end{eqnarray}
With Eq.~(\ref{dt}) the map density corresponding to Eq.~(\ref{dv3}) reads 
\begin{equation}
\varrho_{map}(v) = \frac{2}{E\pi}\frac{v^2}{\sqrt{2E-v^2}}\; .
\label{md3}
\end{equation}
To get the reduced microcanonical probability densities given by
Eqs.~(\ref{dvx3}) and (\ref{dv3}) for the velocity of the particle in
our model we thus have to redefine the baker variable $y_b$.  Conservation
of probability
\begin{equation} 
\varrho_{baker}(y_b)dy_b = \varrho_{map}(v)dv 
\label{tr}
\end{equation}
yields
\begin{equation}
y_b = Y_3(v) = -\frac{v}{\pi
E}\sqrt{2E-v^2}+\frac{2}{\pi}\arcsin\frac{v}{\sqrt{2E}}
\end{equation}
with $0\le v\le \sqrt{2E}\:,\:0\le Y_3(v)\le 1$. The inverse
transformation $v=Y_3^{-1}({y_b})$ exists because $Y_3$ is
monotonous. With $x_b=X(\gamma )=\sin|\gamma|$, we can summarize the
collision rules to
\begin{equation}
(\gamma',v')=(X^{-1},Y_3^{-1})\circ M\circ(X(\gamma),Y_3(v))\; .
\label{coru}
\end{equation}
$k'$ being obtained from energy conservation, $k'=\sqrt{2E-{v'}^2}$.
Fig.~\ref{figd3} shows the probability densities for the three--dimensional system resulting from numerical simulations at 
$E=0.5$. $\varrho_3(\beta)=\varrho_3(\alpha)=1/2\pi$ are uniform
as expected. $\varrho_3(v)$, $\varrho_3(v_x)$ and $\varrho_3(v_y)$ are in
exact agreement with the reduced microcanonical densities, that is,
$\varrho_3(v)$ corresponds to Eq.~(\ref{dv3}) and $\varrho_3(v_x)$,
$\varrho_3(v_y)$ correspond to Eq.~(\ref{dvx3}). Moreover, numerical
simulations show that the trajectory of the particle covers the
Lorentz gas cell uniformly in configuration space.  Beside the
equidistribution of $\varrho_3(\beta)$, this is a further check of the 
ergodic behavior of our system.

\subsection{Arbitrarily many degrees of freedom on the disk}
We have considered an energy transfer between particle and disk for the
case when the disk is equipped with one degree of freedom. We now 
further modify the dynamics by pretending that the disk has
arbitrarily many degrees of freedom $\vec{k}=(k_1,\ldots,k_{d-2})$,
entailing that in total the system has $d$ degrees of freedom. 
We do not deal with the individual components of $\vec{k}$, because the
detailed dynamics of $\vec{k}$ is not relevant for our
purpose. Instead, we consider only the absolute value $|\vec{k}|$.  
The description of our model as a
dynamical system is therefore still based on having only three relevant variables for the
velocities, whereas the corresponding statistical physical situation involves a microcanonical probability density of a $d$-dimensional system, and thus implicitly mimics the situation of having $d$ physical
degrees of freedom.

The consideration of such additional degrees of freedom requires a
modification of the collision rules to get correctly the 
microcanonical probability density corresponding to this $d$--dimensional
system. In particular, we have to redefine the transformation $Y$,
which can be done by the same method as for the
three dimensional system above: First, we calculate the reduced densities
$\varrho_d(v_x)$ and $\varrho_d(v)$ of the $d$-dimensional energy
hypersphere. They are given by Eq.~(\ref{dvx}) and Eq.~(\ref{dv}) in the 
Appendix,
  
\begin{equation}
\varrho_d(v_x) =
 \frac{\Gamma(\frac{d}{2})}{\sqrt\pi\Gamma(\frac{d-1}{2})}
\frac{1}{(2E)^\frac{d-2}{2}}(2E-v_x^2)^\frac{d-3}{2} 
\label{gaus}
\end{equation}
and
\begin{equation}
\varrho_d(v)=\frac{d-2}{(2E)^{\frac{d-2}{2}}}v(2E-v^2)^\frac{d-4}{2}\;.
\label{gausv}
\end{equation}

\begin{figure}[htbp]
\vspace*{-6.cm}
\epsfxsize=14cm
\centerline{\epsfbox{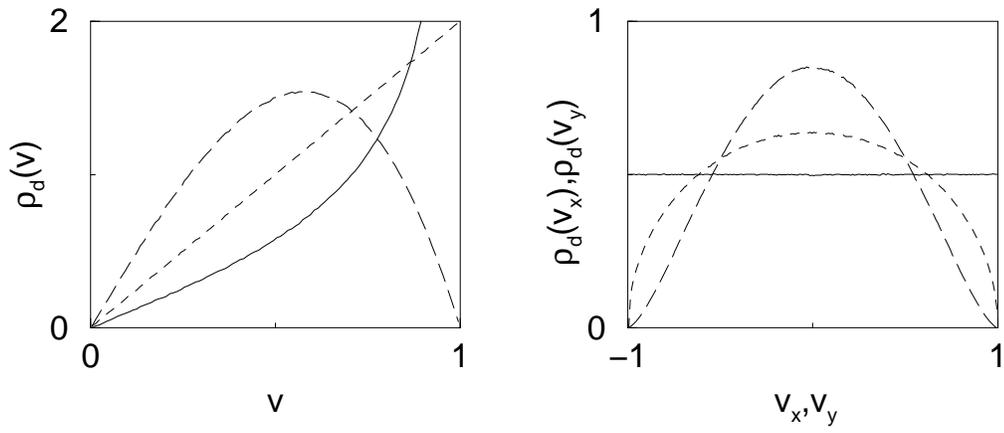}}
\caption{Probability densities for $d=3$ (solid curve), $4$ (dashed curve) and $6$ (long dashed curve) at $E=0.5$}
\label{figd346}
\end{figure}

Inserting Eq.~(\ref{gausv}) into Eq.~(\ref{dt}) leads to the
corresponding map density, and with Eq.~(\ref{tr}) we can calculate
$Y_d$. For arbitrary even $d$ it is given by

\begin{equation}
Y_d = \frac{2\Gamma(\frac{d+1}{2})}{\Gamma(\frac{3}{2})
\Gamma(\frac{d-2}{2})(2E)^\frac{d-1}{2}(d-2)}
\left(-v(2E-v^2)^\frac{d-2}{2}+\sum_{i=0}^\frac{d-2}{2}{\frac{d-2}{2}
\choose i}\frac{(-1)^i}{2i+1}v^{2i+1}(2E)^{\frac{d-2}{2}-i}\right) \: .
\label{dg}
\end{equation}
For all $d$ this expression is monotonous, and thus its inverse always
exists. For odd $d>3$ $Y_d$ can be calculated in the same way, but 
cannot be written down in a closed form.  Inserting Eq.~(\ref{dg}) in
Eq.~(\ref{coru}) yields the full collision rules of our model for
arbitrary even $d$. $|\vec{k'}|$ is obtained from energy conservation.
Fig.~\ref{figd346} shows computer simulation results of $\varrho_d(v)$
and $\varrho_d(v_x)$ for $d=3$, $4$ and $6$. They are in exact agreement
with Eqs.~(\ref{gaus}) and (\ref{gausv}).

Eq.~(\ref{gaus}) was already written by Maxwell and Boltzmann
\cite{Ma1879,Bo09}, however, for their calculation they took a
different starting point: a set of $n$ particles, any particle with
two degrees of freedom, is moving according to Hamilton's equations of
motion. Using properties of the canonical transformation they derived
the reduced microcanonical probability densities Eqs.~(\ref{gaus}) and 
(\ref{gausv}). Notice that starting the calculation of the reduced densities according
to our reasoning, but including also momentum conservation, reduces
the effort of the calculation drastically, see Ref.~\cite{MPT98}.

\subsection{The disk as a thermal reservoir}
We now consider the limit $d\to\infty$.
Using equipartitioning of energy $E=dT/2$ with $k_B=1$,
Eq.~(\ref{dvx}) reduces then to Eq.~(\ref{aivx}),
\begin{equation}
\varrho_\infty(v_x) \ = \ \frac{1}{\sqrt{2\pi T}} e^{-\frac{v_x^2}{2T}} \; ,
\label{ivx}
\end{equation}
which is the canonical distribution for the moving particle. The disk
now acts as a thermal reservoir, yet, the whole system remains
microcanonical. In the same way the limiting form of Eq.~(\ref{dv}), see 
Eq.~(\ref{aiv}), reads
\begin{equation}
 \varrho_\infty(v) \ = \ \frac{1}{T}ve^{-\frac{v^2}{2T}} \; . 
\label{iv}
\end{equation}
As for the transformation $Y_\infty$, it is given by
\begin{equation}
Y_\infty(v) = -\sqrt{\frac{2}{\pi T}}ve^\frac{-v^2}{2T}+\mbox{erf}(\frac{v}{\sqrt{2T}})\;.
\label{iy}
\end{equation}
As before, $Y_\infty$ is monotonous, and thus the inverse $Y_\infty^{-1}$ exists.
Notice that in all these expressions the temperature $T$ appears instead of the
previous total energy of the system and serves as a free parameter.

\begin{figure}[htbp]
\epsfxsize=14cm
\centerline{\epsfbox{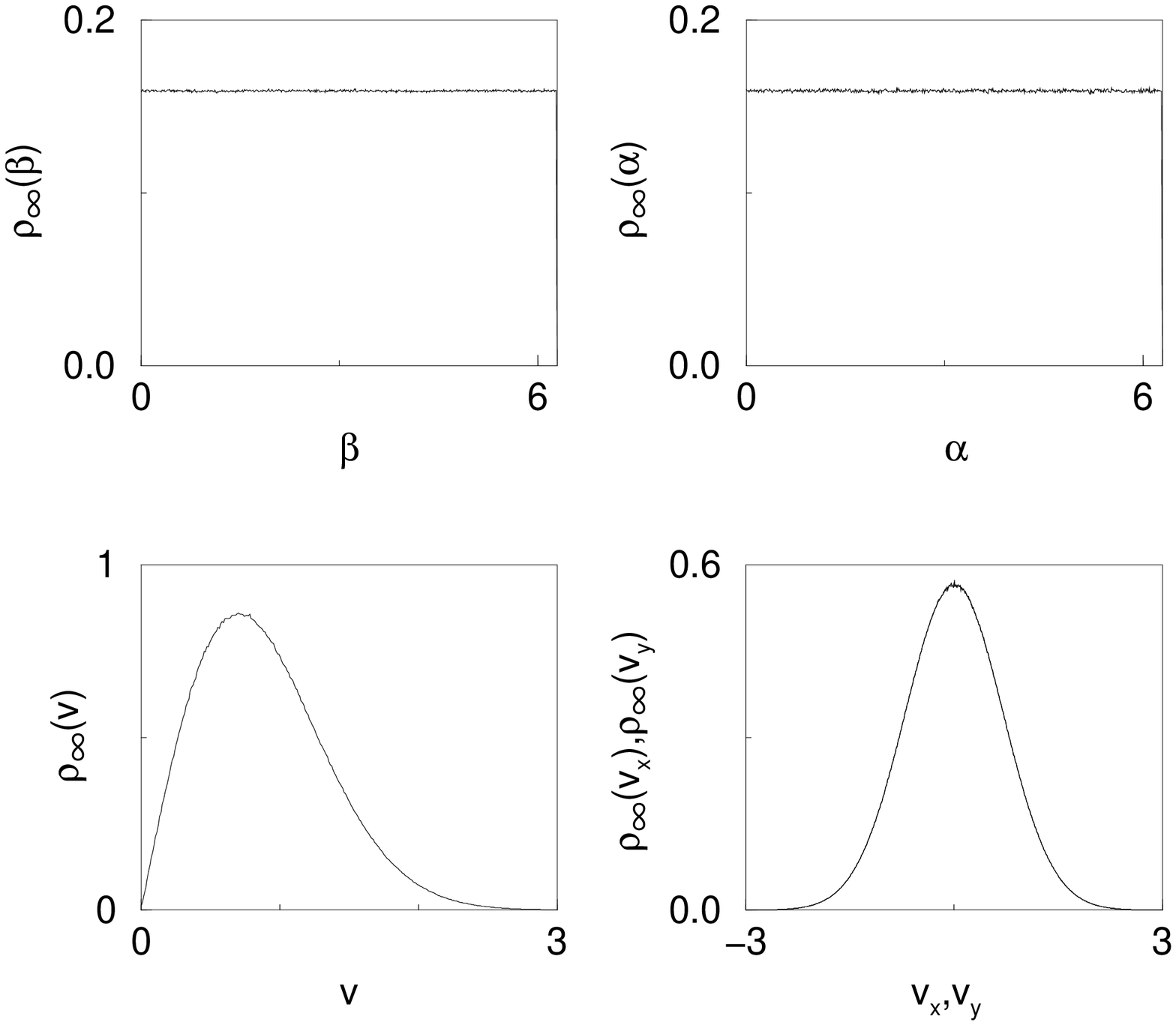}}
\caption{Probability densities for $d\to\infty$ at $T=0.5$}
\label{figuf=0}
\end{figure}

Fig.~\ref{figuf=0} shows the probability densities for $d\to\infty$
resulting from computer simulations: $\varrho_\infty(\beta)$ and
$\varrho_\infty(\alpha)$ are uniform, as expected. $\varrho_\infty(v)$ is in exact agreement with Eq.~(\ref{iv}) and $\varrho_\infty(v_x)$ and $\varrho_\infty(v_y)$ are in exact agreement with Eq.~(\ref{ivx}).

We note that combining Eq.~(\ref{iv}) with the respective equilibrium
distribution for the angle of incidence
$\varrho_\infty(\gamma)=\cos\gamma$ and applying Eq.~(\ref{dt}) yields a
map density $\varrho_{map}(\gamma,v)$ in local polar coordinates which is
identical to the stochastic boundary conditions as given, e.g., 
in Ref.\ \cite{ChLe97}. This relation of our method to stochastic boundary 
conditions is discussed in more detail in Ref.\ \cite{WKN98}, where it has been used as an 
alternative starting point to define the deterministic and time-reversible 
thermostat constructed above.

\section{Nonequilibrium, steady states, and transport under an external electric field} 

In the previous section we have extended the dynamics of a periodic
Lorentz gas in equilibrium by defining new collision rules for the
particle allowing for an energy transfer between particle and disk at
each collision such that the dynamics is deterministic, time
reversible, and yields a microcanonical probability density. In this
section we analyze the behavior of the model under nonequilibrium conditions,
associated with the presence of an external electric field
$\varepsilon$ applied parallel to the $x$--axis. Taking the collision
rules as defined in equilibrium, we study the structure of $\varrho(v)$ in nonequilibrium and the associated transport properties.

\begin{figure}[htbp]
\epsfxsize=10cm
\centerline{\epsfbox{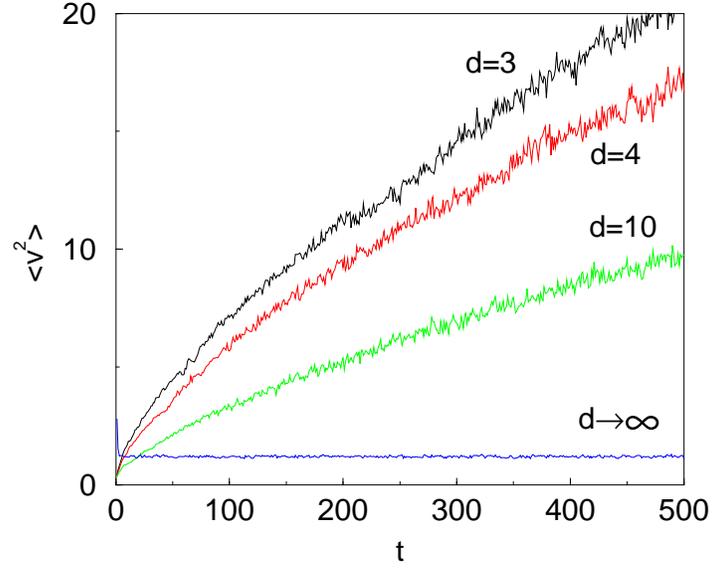}}
\caption{Time series of the ensemble average $<v^2>$  as a function of time $t$ for different d}
\label{ens}
\end{figure}

Fig.~\ref{ens} shows the time evolution of twice the average kinetic energy, 
$<v^2>$, for an ensemble of moving particles and for
different $d's$. As can be seen, the particle energy grows
continuously with time for finite $d$, but fluctuates around a constant mean 
value as $d\to\infty$. This can be understood as
follows. In equilibrium the energy transfer ensures 
equipartitioning of energy between all degrees of freedom. In
the presence of a field the energy of the particle grows during the free
flight, and as a consequence the particle has on the average a surplus
of energy at a collision in comparison to the disk.  The energy
transfer then counteracts this surplus of energy of the
particle, because equipartitioning of energy is still built into the collision rules. Since there is no other source of dissipation in our model
than the transfer of energy onto the disk, the energy of the particle
must eventually grow for finite dimension, while the growth rate
decreases by increasing $d$, because by increasing $d$ more energy can be stored onto the disk. In the limit of $d\to\infty$ we obtain a constant 
average kinetic energy, since then the disk acts as a thermal reservoir with infinitely many degrees of freedom, which means that our system
is thermostated. Still, the system is time reversible. This interesting fact is reflected in the special functional form of the transformation $Y_\infty(v)$ in Eq.~(\ref{iy}).
 
\begin{figure}
\epsfxsize=10cm
\centerline{\epsfbox{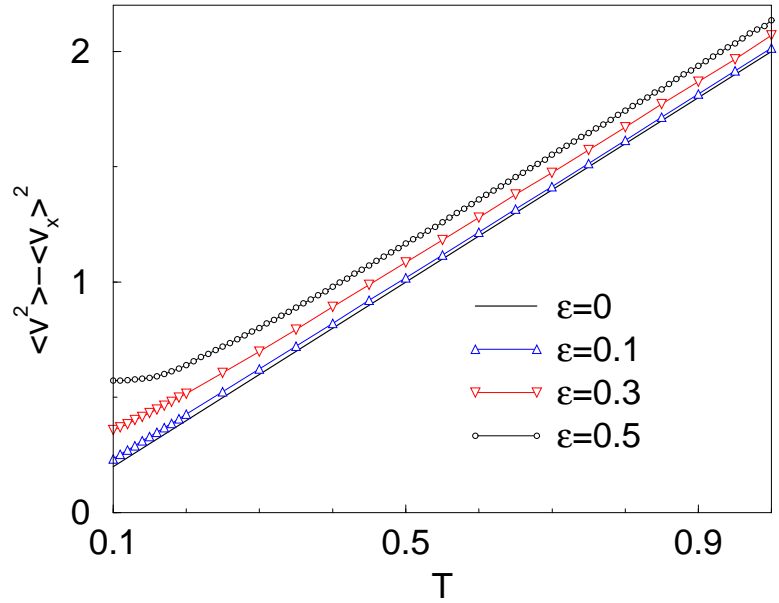}}
\vspace*{0.3cm} 
\caption{Relation between average velocity squared in
the frame moving with the current, $<v^2>-<v_x>^2$, and parametric
temperature $T$ for the infinite dimensional model. Equipartitioning
of energy would imply $<v^2>-<v_x>^2=2T$. The
numerical uncertainty of each point is less the size of the symbols}
\label{kin}
\end{figure}

In the following we investigate the nonequilibrium steady state of the model for $d\to\infty$ in more detail.
The mean kinetic energy in the comoving frame
$<v^2>-<v_x>^2$ of the moving particle in comparison to the one
obtained from the equipartition theorem, $<v^2>-<v_x>^2=2T$, is presented in Fig.~\ref{kin}. For small
$\varepsilon$, or for large temperatures, the curves approach the
equipartitioning values. However, for general $\varepsilon$ and $T$
there is a systematic difference between the measured temperature in
the simulations and the parametric temperature as it appears in the
collision rules. The reason for this difference can be found in our
approach to define the collision rules in an equilibrium situation:
First, Eq.~(\ref{dt}), which is one step in the derivation of the
transformation $Y_\infty$, is based on having a constant particle
velocity between two collisions, which is not correct anymore by
applying an electric field. And second, we always try to transform the
system onto an equilibrium distribution by assuming that this will
still be a good approximation for a system slightly perturbed out of
equilibrium. We note that we do not have any better choice, since we
do not know the correct nonequilibrium distribution for the Lorentz
gas. As a consequence, the parametric temperature of the disk, which
refers to an equilibrium distribution, and the measured kinetic
temperature of the moving particle in nonequilibrium do not agree. One
way out of this problem would be to redefine a proper kinetic
temperature for the reservoir. This can in fact be done by employing
the average value of the variances of the in- and outgoing map
densities, as is discussed in detail in Ref.~\cite{WKN98}. With
respect to such a new temperature definition for the reservoir, we
would expect to have equipartitioning of energy at least for
sufficiently small field strengths when the temperature is
sufficiently high. Fig.~\ref{kin} shows that for low $T$ the
variance approaches finite values not equal to zero for non-vanishing field
strengths. The reason for this behavior is that between collisions the
electric field accelerates the particle, whereas the thermostat acts
only at the collisions. Thus, we may expect that in this region there
will always remain a discrepancy between the measured temperatures of
particle and reservoir.

\begin{figure}[htbp]
\epsfxsize=14cm
\centerline{\epsfbox{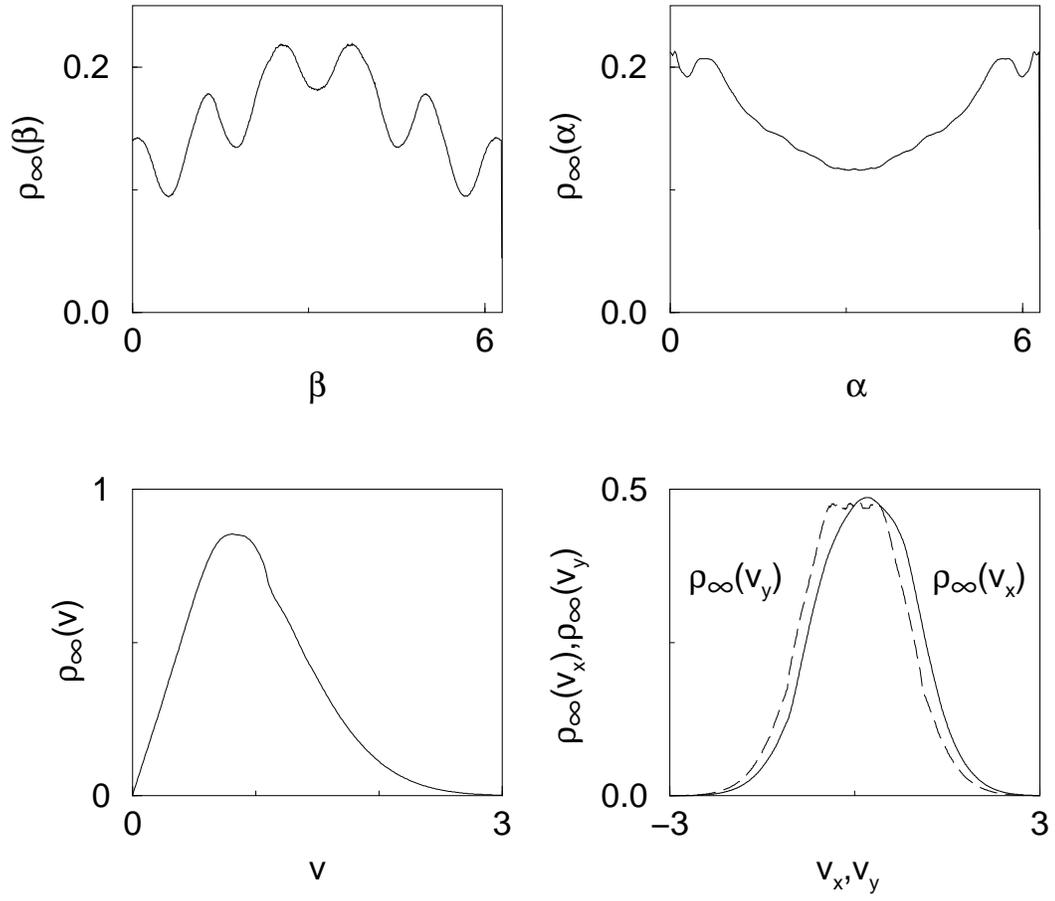}}
\caption{Probability densities for $d\to\infty$, $\varepsilon=0.5$ at $T=0.5$}
\label{figf05}
\end{figure}

Fig.~\ref{figf05} depicts the probability densities for
$d\rightarrow\infty$ and $\varepsilon=0.5$.  The external field leads
to strong deviations from the equilibrium probability densities:
On a very coarse scale, $\varrho_\infty(\beta)$ shows the existence
of a global maximum opposite to the field direction at
$\beta=\pi$. However, there appear six strong pronounced local maxima which are related to the specific geometry of our model. Correspondingly,
$\varrho_\infty(\alpha)$ shows a global maximum parallel to the field
direction at $\alpha=0$, but there also exist four local maxima on a finer
scale. $\varrho_\infty(v)$, $\varrho_\infty(v_x)$ and $\varrho_\infty(v_y)$ are also clearly
modified by the field, while remaining close to the functional form of
the equilibrium probability densities. Especially, $\varrho_\infty(v_x)$ is
shifted to positive $v_x$-values. This indicates the existence of a current 
parallel to the field direction.

\begin{figure}[htbp]
\epsfxsize=14cm
\centerline{\epsfbox{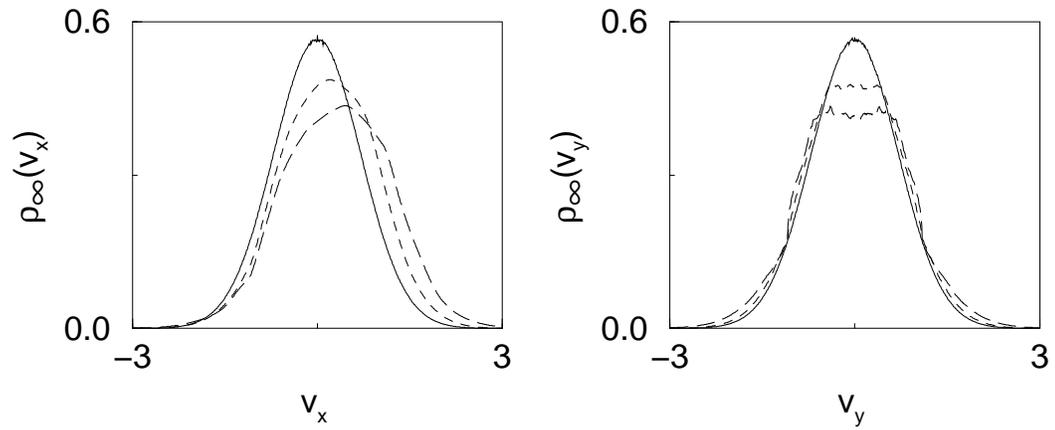}}
\caption{Probability densities for $d\to\infty$ and $\varepsilon=0$ (solid curve), $0.5$ (dashed curve) and $1$ (long dashed curve) at $T=0.5$}
\label{figuf}
\end{figure}

The changes in the probability densities by increasing the field
strength in reference to the equilibrium solutions are presented 
in Fig.~\ref{figuf}. The mean value of $\varrho_\infty(v_x)$
grows with $\varepsilon$, whereas $\varrho_\infty(v_y)$ remains symmetric
around $0$. 

\begin{figure}
\epsfxsize=10cm
\centerline{\epsfbox{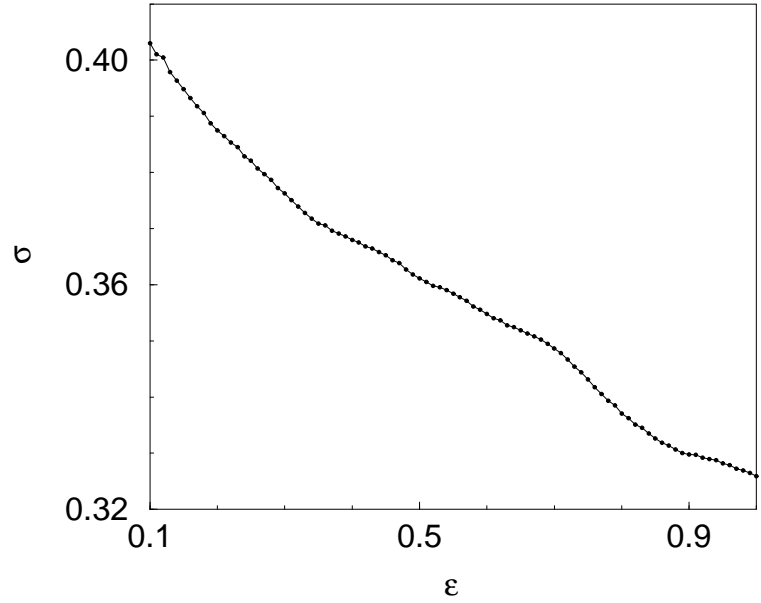}}
\vspace*{0.3cm} 
\caption{Conductivity $\sigma(\varepsilon)$ as it varies with field
strength $\varepsilon$ for $T=0.5$. The curve consists of 90 data points, the
numerical uncertainty of each point is less the size of the symbols}
\label{cond} 
\end{figure}

The conductivity $\sigma=<v_x>/\varepsilon$, shown in Fig.~\ref{cond}, 
is a globally decreasing function of 
$\varepsilon$. This demonstrates that for the field strengths considered in 
the figure we are
already in a highly nonlinear regime. We furthermore note that there
exist irregularities in $\sigma$ on a finer scale, which are
beyond our numerical error estimates and indicate a deviation of
$\sigma$ from a simple functional dependence on $\varepsilon$. This
curve should qualitatively be compared to the conductivity as obtained
for the Gaussian thermostated periodic Lorentz gas\footnote{Note that in Refs.~\cite{MH87,DeGP95} the angle between the field direction and
  the x--achses has been chosen equal to $\frac{\pi}{6}$.}
\cite{MH87,DeGP95,LNRM95}: Fig.~2 in 
Ref.~\cite{LNRM95} shows a globally decreasing conductivity on
a coarse scale as well, however, its irregularities on a fine scale
are much more pronounced and clearly nonmonotonous in $\varepsilon$.
Whether the irregularities in our Fig.\ \ref{cond} are in fact
monotonous or nonmonotonous on a finer scale cannot be decided on the basis
of our numerical data. Unfortunately, it is not clear how to compare
the conductivities of these two models quantitatively, since the
choice of temperature in the Gaussian model is somewhat ambiguous by a
factor of two \cite{Ch1,Ch2}.

To understand why no linear response is seen in
Fig.~\ref{cond}, we attempt an estimate of the expected range of validity of 
the
linear response regime by applying the simple heuristic argument
suggested, e.g., in Refs.\ \cite{Ch2,Do99} and further references
therein. The reasoning given in these references may actually be
considered as the standard response to the famous van Kampen objection
against linear response \cite{vK71}, since it amounts to properly
modifying his original argument. For the periodic Lorentz gas it is stipulated that for having linear response it is sufficient
for the field strength to fulfill $\varepsilon\ll 2/\tau_{coll}^2$,
where $\tau_{coll}$ is the average mean-free time between collisions.
We have computed this bound for the Gaussian thermostated Lorentz gas
as well as for our model. For the Gaussian thermostat, $\tau_{coll}$
has already been obtained from computer simulations at different
densities of scatterers in Ref.\
\cite{DeGP95}.  For the density corresponding to the gap size
$w=0.2361$ the upper bound for linear response is then approximately
$\varepsilon\ll 5.6$. However, numerical results for the conductivity
indicate no linear response down to $\varepsilon \simeq 0.05$
\cite{LNRM95}.  For a higher density corresponding to $w=0.129$, the conductivity in Fig.\
9 of Ref.\ \cite{DeGP95} and the respective data values {\em may} be
compatible with a linear response-like behavior at most up to a field strength
of approximately $\varepsilon\approx 0.9$. Here, an upper bound for linear
response yields $\varepsilon\ll 13$. Finally, for $w=0.0076$ the
respective upper bound is $\varepsilon\ll 70$. The corresponding
conductivity appears to be well-behaved up to $\varepsilon\simeq 2.5$,
although for this very high density the numerical values are not very
precise. For our model, we calculated as an upper bound $\varepsilon\ll
4.45$, however, down to a field strength of $\varepsilon=0.05$ we
cannot detect any linear response.

We conclude that in the light of the numerical simulations the bound for
linear response discussed above does not appear to serve as a very reliable
reference for the driven periodic Lorentz gas. Other restrictions,
which may make this bound more precise, have been discussed
in Ref.\ \cite{Ch2}, without leading, however, to an improved
quantitative formulation. Thus, although the existence of a linear
response has been proven for the Gaussian thermostated
periodic Lorentz gas with an external field in Refs.~\cite{Ch1,Ch2}, having a
reasonable estimate for the range of validity of linear response in
this system remains, to our knowledge, an open question. In
particular, we note that up to now for smaller densities like
$w=0.2361$ such a regime has not been detected in computer simulations.
We furthermore note that analogous problems, namely the existence of
irregularities on a fine scale in parameter-dependent deterministic
transport coefficients as well as apparently the complete lack of a
linear response regime have already been encountered in more simple
low-dimensional models of chaotic transport. These are so-called
multibaker maps, which are believed to represent certain essential dynamical features of Lorentz gas models \cite{Gasp,GaKl,BTV98}, and associated one-dimensional maps. The currents in such systems have been
computed by numerically exact methods with respect to varying a 
bias, and they turned out to be fractal functions in the bias
\cite{Groe95,mapg,mapb}, implying also the non-existence of a
regime of linear response down to extremely small field strengths. 
Such properties may be related to the low dimensionality of these systems, which is shared by the periodic Lorentz gas.

\begin{figure}
\epsfxsize=10cm
\centerline{\epsfbox{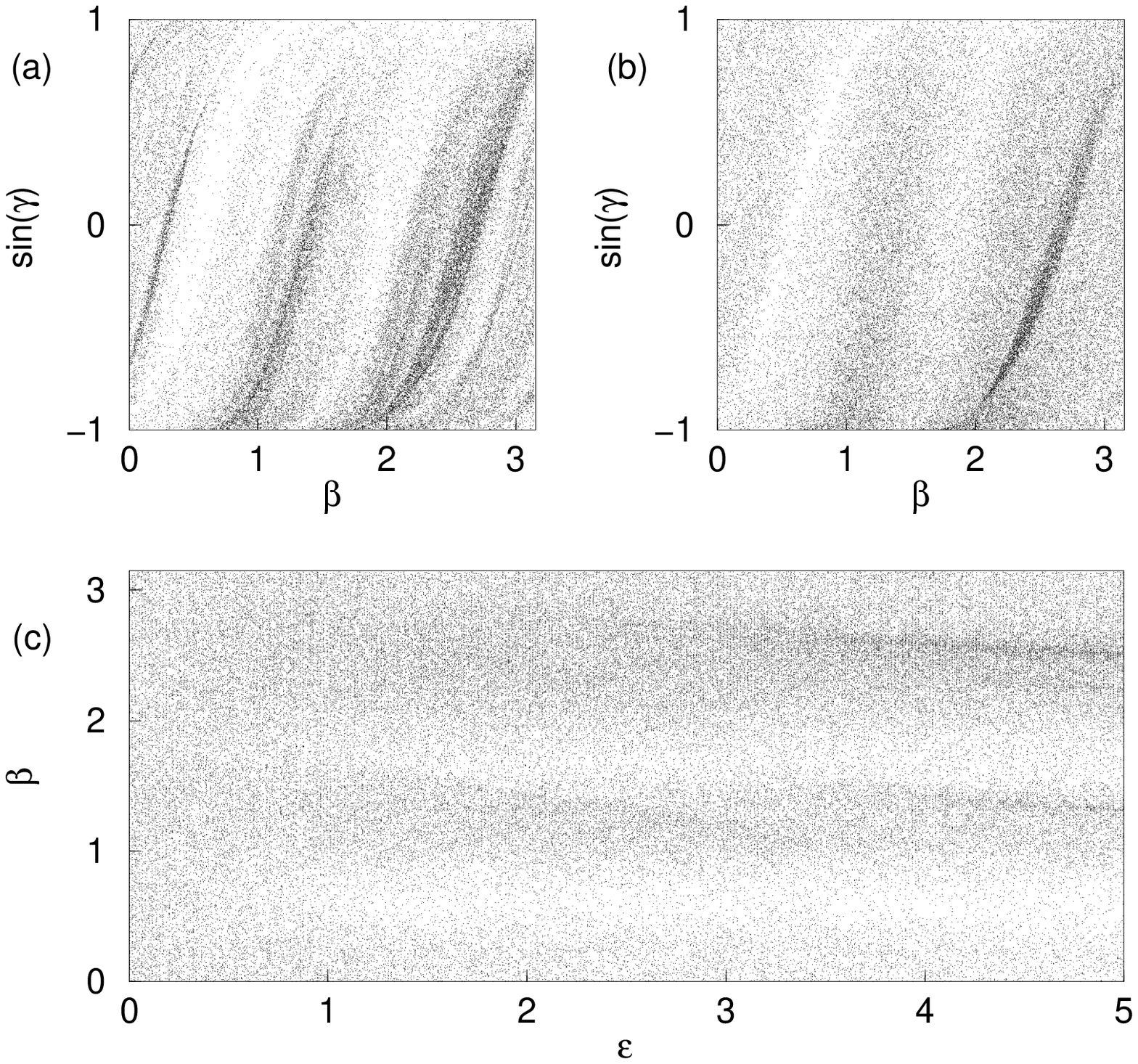}}
\vspace*{0.3cm} 
\caption{(a),(b) Poincar\'e section of $(\beta,\sin\gamma)$ defined in Fig.\ref{cell} at the moment of the collision for field strength
$\varepsilon=1$. In (a) our model with the baker map has been
used, in (b) the baker has been replaced by a random number
generator. (c) Poincar\'e section of $\beta$ at the moment of the
collision for varying field strength $\varepsilon$.(a).(b),(c) with $T=0.5$.}
\label{bifu}
\end{figure}

Fig.~\ref{bifu} (a) shows the Poincar\'e section of
$(\sin(\gamma),\beta)$ at the moment of the collision obtained for our model 
with the baker map. It displays a highly variable phase space density exhibiting a structure which is
qualitatively analogous to the one of the fractal attractor found for
the Gaussian thermostated Lorentz gas
\cite{MH87,HoMo89,HooverPosch98,HooverPosch98a,DeGP95}. The existence
of such an attractor for our model is due to the fact that the phase space variables in Fig.\ \ref{bifu} (a) only reflect the dissipative dynamics of the moving particle, whereas the corresponding
complementary dynamics of the reservoir is completely left out. However, the dynamics of the single moving particle, which represents here a subsystem projected out of
the full system (particle plus reservoir), is certainly not phase space
volume preserving, because there is an average energy transfer to the
reservoir. This does not contradict the existence of volume preservation, and
thus a related Hamiltonian character, of the {\em full} system
(particle plus reservoir) if one takes the complete dynamics of the full
system appropriately into account. This reasoning can be made more precise by
computing the Jacobian determinant for the full system (particle plus
reservoir) at a collision by systematically varying the number of
degrees of freedom associated to the reservoir \cite{RKj}.

Figure \ref{bifu} (b) shows the Poincar\'e section for a
modification of our model where we have replaced the baker map by a
random number generator, thus modeling in a way stochastic 
boundary conditions. This stochastic model apparently leads to a
smooth nonuniform density.
The existence of smooth versus singular measures in
thermostated dynamical systems like the driven Lorentz gas has been
vividly discussed in the recent literature
\cite{GKI85,HooverPosch98a,EyLe92,PoschHoover98}: As 
pointed out in the introduction of this paper, the existence of singular
nonequilibrium measures is an essential feature of Gaussian and
Nos\'e-Hoover thermostated systems.  On the other hand, it has been
proven that for a specific system in a nonequilibrium
situation created by stochastic boundaries the corresponding
measure is smooth \cite{GKI85}, and it has been argued that this should 
be considered as a general feature of systems which are thermostated
by stochastic boundaries \cite{EyLe92}.
Our results of Figs.~\ref{bifu} (a) and (b) suggest that for the
periodic Lorentz gas there is a clear distinction between
deterministic boundaries producing a singular measure, and stochastic
boundaries creating a smooth measure. We note that in Ref.\
\cite{HooverPosch98a} a combination of deterministic and stochastic
boundaries has been applied to the driven periodic Lorentz gas,
leading to the observation that the fractal attractor is apparently
very stable with respect to the random perturbations induced by the
stochastic part of the boundaries. In Ref.\ \cite{PoschHoover98}
computer simulations appear to indicate the existence of a slightly
singular measure for a system with stochastic boundaries. How the
nonequilibrium stationary measure looks like for a general
many-particle systems with stochastic boundaries remains thus an open
question up to now. Our approach of thermostating may help to bridge
the gap, because we can alternatively produce deterministic or
stochastic boundaries simply by replacing the reversible deterministic
map by a random number generator, thereby either preserving or
completely destroying any dynamical correlations induced by the
boundaries.

Fig.~\ref{bifu} (c) shows the Poincar\'e plot for
$\beta=f(\varepsilon)$ at $T=0.5$. For small $\varepsilon$ we observe
a uniform phase space density which is in agreement with the histogram
$\varrho_\infty(\beta)$ in Fig.~\ref{figuf=0}. By increasing
$\varepsilon$ the density shows several maxima but remains phase space
filling. This behavior reflects the dynamics depicted in the histogram
$\varrho_\infty(\beta)$ of Fig.~\ref{figf05}. We could neither detect
a contraction of the attractor onto periodic orbits, nor a breakdown
of ergodicity for higher field strengths, or eventually the 
existence of a so-called creeping orbit for very high field
strengths\footnote{If the alternation of forward and backward baker as introduced in section \ref{zb} is not fine enough one observes periodic windows, which disappear in the limit of infinitely fine steps.}. 
Similar results have been obtained for other choices of
Poincar\'e sections in phase space. 
 This is in contrast to what has been found for the Gaussian isokinetic 
thermostat
\cite{MoDe98,MH87,DettmannMorriss96,LRM94,LNRM95}, suggesting that the
complicated scenario found in the Gaussian model is rather a property
induced specifically by the Gaussian thermostating mechanism.

\section{Conclusions} 

We have presented a new deterministic thermostating mechanism for the
periodic Lorentz gas. Our mechanism is based on modeling the energy
transfer between the particle and the disk at each collision instead
of using a momentum--dependent friction coefficient. We have defined
the collision rules for the particle such that they are time
reversible, deterministic, and yield a microcanonical probability
density in equilibrium. 
We have shown how our mechanism could be
applied for a disk equipped with arbitrarily many degrees of freedom.
We investigated the behavior of the model under nonequlibrium conditions 
by applying an external electric field. In the limit
$d\rightarrow\infty$, where the disk acts as a thermal reservoir, we
obtained a nonequilibrium steady state with a constant average
kinetic energy of the particle implying that we have successfully
thermostated our system.

With respect to the parametric temperature contained in our collision
rules the dynamics of our model nearly fulfills equipartitioning of
energy for small field strengths, however, by increasing the field
strength the deviation from equipartitioning grows. This appears to be
a problem of a definition of temperature in nonequilibrium and
could probably be avoided by using an appropriately defined temperature
for the disk.

The results obtained from our model in nonequilibrium
have been compared to the results of the Gaussian thermostated periodic Lorentz gas. It
turned out that the attractor associated to the model is qualitatively 
similar to
the fractal attractor of the Gaussian thermostated periodic Lorentz
gas. However, we could not find any breakdown of ergodicity or a 
collaps of the attractor onto periodic orbits for higher field
strengths in our model. In both models the conductivity is a nonlinear
decreasing function with increasing field strength with more or less
pronounced irregularities on a finer scale.

Our thermostating mechanism yields the canonical probability density
for the moving particle in equilibrium and keeps the kinetic energy of
the particle constant on average in nonequilibrium. 
This is similar to what is achieved by applying a Nos\'e--Hoover thermostat.
However, to our knowledge a Nos\'e--Hoover thermostated Lorentz gas has not yet been studied so far. 
Work along these lines and comparison with the results of our model is currently under investigation \cite{RKHN99}.
 
We note that the thermostating mechanism presented in this work has
recently been successfully applied to simulate a heat and shear flow of
an interacting many-particle system of hard disks \cite{WKN98}. As a
main result, it has been found that in general there exists no
identity between phase space contraction and entropy production in
this system if it is thermostated by our method. We would expect the
same result to hold for the thermostated Lorentz gas as described in
this paper. It would also be interesting to investigate the validity
of fluctuation theorems \cite{GaCo95a,LeSp98} for our Lorentz gas model.

Further work along the lines of this paper should also aim at calculating
the Lyapunov exponents, using for instance the method of Dellago and
Posch
\cite{DeGP95,DePo95,DePH96}. This would enable to check for the
validity of the expression of the conductivity in terms of the sum of
Lyapunov exponents, as obtained for conventionally thermostated
systems
\cite{MH87,PoHo88,Ch1,Ch2,ECM,Vanc,BarEC,DeGP95}. Moreover, it could
be investigated whether there exists a conjugate pairing rule for the
Lyapunov exponents in our model \cite{MoDe98,ECM}. 
In particular, the knowledge of the Lyapunov exponents would allow an
estimation of the fractal dimension of the attractor.

Finally, collision processes in granular media might be advantageously
modeled by applying our formalism of energy transfer. This could be
done by associating finitely many degrees of freedom to a scatterer
according to our method, thus rendering the collision process
inelastic. This would provide an interesting
alternative to describing collision processes by velocity dependent
restitution coefficients, as it has been done previously
\cite{BSHP96,PoSch97,SchPo98,GiZi96,AGZ98}.

{\bf Acknowledgements:} We are indebted to P.Gaspard for many
important hints. Helpful discussions with W.G.Hoover are gratefully
acknowledged. K.R.\ thanks the European Commisssion for a TMR grant
under contract no.\ ERBFMBICT96-1193, and R.K.\ thanks the DFG and the  
European Commission for financial support.
      
\begin{appendix}
\section{Calculation of the reduced densities}
In this section we calculate the reduced, or projected, densities for
one and two degrees of freedom of a $d$--dimensional microcanonical
system.  We use the following notation: general momentum space
coordinates are denoted by $x_1,\ldots,x_d$ or by $x$, and $w_i$
is the absolute value of a vector with $i\:,\:i\in N$, degrees of
freedom.

The probability density of a $d$--dimensional microcanonical system
is an equidistribution on a $d$--dimensional hypersphere,
\begin{equation}
\varrho_d(x_1,\ldots,x_d)=\frac{\Gamma(\frac{d}{2})}{2\pi^{\frac{d}{2}}(2E)^\frac{d-1}{2}}\delta(2E-x_1^2-\ldots
-x_d^2)\; .
\label{mic}
\end{equation}

With the definition of $d$-dimensional spherical coordinates
\begin{eqnarray}
x_1 &=& \sqrt{2E}\sin\psi_{d-1}\sin\psi_{d-2}\ldots\sin\psi_2\sin\psi_1 \nonumber\\
x_2 &=& \sqrt{2E}\sin\psi_{d-1}\sin\psi_{d-2}\ldots\sin\psi_2\cos\psi_1 \nonumber\\
x_3 &=& \sqrt{2E}\sin\psi_{d-1}\sin\psi_{d-2}\ldots\cos\psi_2 \nonumber\\
\ldots \nonumber\\
x_{d-1} &=& \sqrt{2E}\sin\psi_{d-1}\cos\psi_{d-2} \nonumber\\
x_d &=& \sqrt{2E}\cos\psi_{d-1}\;, 
\label{spheco}
\end{eqnarray}
where $\psi_1\in[0,2\pi)$ and $\psi_n\in[0,\pi), n\geq2$,
we can transform Eq.~(\ref{mic}) onto these coordinates according to
\begin{equation}
\varrho_d(\sqrt{2E},\psi_1,\ldots,\psi_{d-1})=J_d(x_1,\ldots,x_d,\sqrt{2E},\psi_1,\ldots,\psi_{d-1})\varrho_d(x_1,\ldots,x_d)\; ,
\label{trafo}
\end{equation}
where $J_d=|(dx_1\ldots dx_d)/(d\sqrt{2E}d\psi_{d-1}\ldots d\psi_1)|$ is the Jacobian determinant. 
For $J_d$, $d\geq3$, we obtain the following
recursion relation, 
\begin{equation}
J_d=\left|
\begin{array}{*{5}{c}}
\sin\psi_{d-1}\alpha_{11} & \sqrt{2E}\cos\psi_{d-1}\alpha_{11} & \sin\psi_{d-1}\alpha_{12} & \ldots & \sin\psi_{d-1}\alpha_{1(d-1)} \\
\sin\psi_{d-1}\alpha_{21} & \sqrt{2E}\cos\psi_{d-1}\alpha_{21} & \sin\psi_{d-1}\alpha_{22}& \ldots & \sin\psi_{d-1}\alpha_{2(d-1)} \\
\multicolumn{5}{c}{\dotfill} \\
\sin\psi_{d-1}\alpha_{(d-1)1} & \sqrt{2E}\cos\psi_{d-1}\alpha_{(d-1)1} & \sin\psi_{d-1}\alpha_{(d-1)2} & \ldots & \sin\psi_{d-1}\alpha_{(d-1)(d-1)} \\
\cos\psi_{d-1} & -\sqrt{2E}\sin\psi_{d-1} & 0 & \ldots & 0
\end{array}
\right|\; ,\nonumber
\end{equation}
where $\alpha_{jk}$ are the matrix elements of the Jacobian $J_{d-1}$
for $d-1$ dimensions. An expansion after the $d$--th row yields
\begin{equation}
J_d=(\cos \psi_{d-1}|\Omega_1|+\sqrt{2E}\sin\psi_{d-1}|\Omega_2|)\; .\nonumber
\end{equation}
Any column in the matrices $\Omega_1$ and $\Omega_2$ is equivalent to
the corresponding column in $J_{d-1}$ multiplied by a factor. In the
first column of $\Omega_1$ (derivative after $\psi_{d-1}$) the factor
is $\sqrt{2E}\cos\psi_{d-1}$, in the other columns of $\Omega_1$ the
factor is $\sin \psi_{d-1}$.  In all columns of $\Omega_2$ the factor
is $\sin \psi_{d-1}$.  Factoring out leads to
\begin{equation}
J_d=\sqrt{2E}\sin^{d-2}\psi_{d-1}J_{d-1}\;. \nonumber
\end{equation}
Proceeding for the Jacobians $J_{d-1} \ldots J_2$ in the same way
gives
\begin{equation}
J_d=(2E)^\frac{d-1}{2}\prod^{d-1}_{n=2}\sin^{n-1} \psi_n\;.
\label{jd}
\end{equation}

Inserting Eq.~(\ref{jd}) into Eq.~(\ref{trafo}) and taking into account that the radius appearing in the spherical coordinates is a constant equal to $\sqrt{2E}$, according to $\delta (2E-x_1^2-\ldots -x_d^2)\equiv 1$ leads to the microcanonical probability density in spherical coordinates
\begin{equation}
\varrho_d(\psi_1,\ldots,\psi_{d-1})=
\frac{\Gamma(\frac{d}{2})}{2\pi^{\frac{d}{2}}}\prod^{d-1}_{n=2}\sin^{n-1} \psi_n\;.
\label{dspher}
\end{equation}

In Eq.~(\ref{dspher}) the angles $\psi_n$ appear in product form. This allows to infer from this equation directly the probability density for the angle 
$\psi_n$,
\begin{equation}
\varrho(\psi_n)=N\sin^{n-1}\psi_n\; .\nonumber
\end{equation}
Normalizing $\varrho(\psi_n)$, $n\ge2$, leads to
\begin{equation}
N=\frac{\Gamma(\frac{n+1}{2})}{\sqrt\pi\Gamma(\frac{n}{2})} \nonumber
\end{equation}
and $\varrho(\psi_1)=1/2\pi$. Now we can calculate the probability density for a
variable representing exactly one degree of freedom by using the last line of Eq.~(\ref{spheco}),
\begin{eqnarray}
\varrho_d(x_d) &=& \varrho(\psi_{d-1})|\frac{d \psi_{d-1}}{dx_d}| \nonumber\\
\varrho_d(x_d) &=& 
\frac{\Gamma(\frac{d}{2})}{\sqrt\pi\Gamma(\frac{d-1}{2})}
\frac{1}{(2E)^\frac{d-2}{2}}(2E-x_d^2)^{\frac{d-3}{2}}\; . 
\label{dvx}
\end{eqnarray}
Eq.~(\ref{dvx}) is valid for all $x_n, 1\leq n\leq d$. To simplify the notation we replace the variable $x_d$ by $x$ in the following. 
Inserting equipartitioning of energy $E = Td/2$ 
into Eq.~(\ref{dvx}) and performing the limit $d\rightarrow\infty$ leads to 
\begin{equation}
\lim_{d\to\infty}(1-\frac{x^2}{Td})^{\frac{d-3}{2}} =
e^{-\frac{x^2}{2T}} \nonumber
\end{equation}
for the last term and, by applying Stirlings formula, to
\begin{equation}
\lim_{d\to\infty}\frac{\Gamma(\frac{d}{2})}{\Gamma(\frac{d-1}{2})}=
\sqrt\frac{d}{2} \nonumber
\end{equation}
for the prefactor. Finally, we get
\begin{equation}
\varrho_\infty(x) = \lim_{d\to\infty} \varrho_d(x) = \frac{1}{\sqrt{2\pi T}}
e^{-\frac{x^2}{2T}}\;.
\label{aivx}
\end{equation}

The probability density $\varrho_d(w_{d-1})$ with $w_{d-1}=\sqrt{x_1^2+x_2^2+\ldots+x_{d-1}^2}=\sqrt{2E-x_d^2}=\sqrt{2E}\sin\psi_{d-1}$ can be calculated straightforward to
\begin{eqnarray}
\varrho_d(w_{d-1})&=&2\varrho(\psi_{d-1})|\frac{d\psi_{d-1}}{dw_{d-1}}| \label{wd1}\\
\varrho_d(w_{d-1})&=&\frac{\Gamma(\frac{d}{2})}{\sqrt\pi\Gamma(\frac{d-1}{2})} 
\frac{2}{(2E)^\frac{d-2}{2}}\frac{w_{d-1}^{d-2}}{\sqrt{2E-w_{d-1}^2}}\; .\nonumber
\end{eqnarray}
Note that the angles $\psi_{d-1}, 0\leq \psi_{d-1} < \pi /2$ and $\pi /2<\pi-\psi_{d-1}\leq \pi$  are mapped onto the same value $w_{d-1}$, $w_{d-1}\geq 0$, which leads to the factor of two in Eq.~(\ref{wd1}).

We proceed by considering the variable 
$w_{d-2}=\sqrt{x_1^2+x_2^2+\ldots+x_{d-2}^2}=w_{d-1}\sin\psi_{d-2}$.
Together with the variable $x_{d-1}=w_{d-1}\cos\psi_{d-2}$ of Eq.~(\ref{spheco}) a two--dimensional set of variables $(w_{d-2},x_{d-1})$ can be defined, where $(w_{d-1},\psi_{d-2})$ are the corresponding polar coordinates. On these grounds the probability density $\varrho_d(w_{d-2})$ can be calculated to
\begin{eqnarray}
\varrho_d(x_{d-1},w_{d-2})&=&\varrho_d(w_{d-1})\varrho(\psi_{d-2})|\frac{\partial(w_{d-1},\psi_{d-2})}{\partial(x_{d-1},w_{d-2})}|\nonumber\\
 &=&\frac{d-2}{\pi(2E)^{\frac{d-2}{2}}}\frac{w_{d-2}^{d-3}}{\sqrt{2E-w_{d-2}^2-x_{d-1}^2}} \;.\nonumber
\end{eqnarray}
Integration over $x_{d-1}$ yields
\begin{equation}
\varrho_d(w_{d-2})=\frac{d-2}{(2E)^{\frac{d-2}{2}}}w_{d-2}^{d-3}\;.
\end{equation}
With $w_2=\sqrt{2E-w_{d-2}^2}$ we get the probability density for two
degrees of freedom $\varrho_d(w_2)$,
\begin{equation}
\varrho_d(w_2)=\frac{d-2}{(2E)^{\frac{d-2}{2}}}w_2(2E-w_2^2)^\frac{d-4}{2}\;.
\label{dv}
\end{equation}
Inserting $E=Td/2$ and performing the limit
$d\rightarrow\infty$ yields
\begin{equation}
\varrho_{\infty}(w_2)=\frac{w_2}{T}e^{-\frac{w_2^2}{2T}}\;.
\label{aiv}
\end{equation}

\end{appendix}

\newpage

\vspace*{0.5cm}
\vspace{0.5cm}

\vspace{0.5cm}

\vspace*{1.0cm}
\vspace{0.5cm}

\vspace{0.5cm}

\vspace{0.5cm}

\vspace*{-6cm}

\vspace*{-1.0cm}

\begin{thebibliography}{10}

\bibitem{HoAs75}
W.~G. Hoover and W.~T. Ashurst,  in {\em Theoretical Chemistry}, edited by H.
  Eyring and D. Henderson (Academic, New York, 1975), Vol.~1.

\bibitem{HG83}
J.~M. Haile and S. Gupta, J. Chem. Phys. {\bf 79},  3067  (1983).

\bibitem{AT87}
M.~D. Allen and D.~J. Tildesley, {\em Computer simulation of liquids}
  (Clarendon Press, Oxford, 1987).

\bibitem{EvMo90}
D.~J. Evans and G.~P. Morriss, {\em Statistical Mechanics of Nonequilibrium
  Liquids} (Academic Press, London, 1990).

\bibitem{Hoov91}
W.~G. Hoover, {\em Computational statistical mechanics} (Elsevier, Amsterdam,
  1991).

\bibitem{Hess96}
S. Hess,  in {\em Computational physics}, edited by K. Hoffmann and M.
  Schreiber (Springer, Berlin, 1996), pp.\ 268--293.

\bibitem{MoDe98}
G.~P. Morriss and C.~P. Dettmann, Chaos {\bf 8},  321  (1998).

\bibitem{HLM82}
W.~G. Hoover, A.~J.~C. Ladd, and B. Moran, Phys. Rev. Lett. {\bf 48},  1818
  (1982).

\bibitem{Ev83}
D.~J. Evans, J. Chem. Phys. {\bf 78},  3297  (1983).

\bibitem{EH83}
D.~J. Evans {\it et~al.}, Phys. Rev. A {\bf 28},  1016  (1983).

\bibitem{Nose84a}
S. Nos\'e, Mol. Phys. {\bf 52},  255  (1984).

\bibitem{Nose84b}
S. Nos\'e, J. Chem. Phys. {\bf 81},  511  (1984).

\bibitem{Hoov85}
W.~G. Hoover, Phys. Rev. A {\bf 31},  1695  (1985).

\bibitem{LeSp78}
J.~L. Lebowitz and H. Spohn, J. Stat. Phys. {\bf 19},  633  (1978).

\bibitem{TCG82}
A. Tenenbaum, G. Ciccotti, and R. Gallico, Phys. Rev. A {\bf 25},  2778
  (1982).

\bibitem{GKI85}
S. Goldstein, C. Kipnis, and N. Ianiro, J. Stat. Phys. {\bf 41},  915  (1985).

\bibitem{KRN98}
R. Klages, K. Rateitschak, and G. Nicolis, preprint chao-dyn/9812021 (1998).

\bibitem{EvHo85}
D.~J. Evans and B.~L. Holian, Phys. Rev. A {\bf 83},  4069  (1985).

\bibitem{MH87}
B. Moran and W.~G. Hoover, J. Stat. Phys. {\bf 48},  709  (1987).

\bibitem{HHP87}
B.~L. Holian, W.~G. Hoover, and H.~A. Posch, Phys. Rev. Lett. {\bf 59},  10
  (1987).

\bibitem{Hoov88}
W.~G. Hoover, Phys. Rev. A {\bf 37},  252  (1988).

\bibitem{Morr87}
G.~P. Morriss, Phys. Lett. A {\bf 122},  236  (1987).

\bibitem{Morr89}
G.~P. Morriss, Phys. Lett. A {\bf 134},  307  (1989).

\bibitem{HoMo89}
W.~G. Hoover and B. Moran, Phys. Rev. A {\bf 40},  5319  (1989).

\bibitem{Mo89a}
G.~P. Morriss, Phys. Rev. A {\bf 39},  4811  (1989).

\bibitem{HooverPosch98}
W.~G. Hoover and H.~A. Posch, Chaos {\bf 8},  366  (1998).

\bibitem{HooverPosch98a}
W.~G. Hoover and H.~A. Posch, Phys. Lett. A {\bf 246},  247  (1998).

\bibitem{EyLe92}
G. Eyink and J. Lebowitz,  in {\em Microscopic simulations of complex
  hydrodynamic phenomena}, Vol.~292 of {\em NATO ASI Series B: Physics}, edited
  by M. Mareschal and B.~L. Holian (Plenum Press, New York, 1992), pp.\
  323--326.

\bibitem{PoHo87}
H.~A. Posch and W.~G. Hoover, Phys. Lett. A {\bf 123},  227  (1987).

\bibitem{PoHo88}
H.~A. Posch and W.~G. Hoover, Phys. Rev. A {\bf 38},  473  (1988).

\bibitem{PH89}
H.~A. Posch and W.~G. Hoover, Phys. Rev. A {\bf 39},  2175  (1989).

\bibitem{Ch1}
N.~L. Chernov, C.~L. Eyink, J.~L. Lebowitz, and Y.~G. Sinai, Phys. Rev. Lett.
  {\bf 70},  2209  (1993).

\bibitem{Ch2}
N.~L. Chernov, C.~L. Eyink, J.~L. Lebowitz, and Y.~G. Sinai, Comm. Math. Phys.
  {\bf 154},  569  (1993).

\bibitem{ChLe95}
N.~I. Chernov and J.~L. Lebowitz, Phys. Rev. Lett. {\bf 75},  2831  (1995).

\bibitem{TeVB96}
T. Tel, J. Vollmer, and W. Breymann, Europhys. Lett. {\bf 35},  659  (1996).

\bibitem{VTB97}
J. Vollmer, T. T\'el, and W. Breymann, Phys. Rev. Lett. {\bf 79},  2759
  (1997).

\bibitem{ChLe97}
N.~I. Chernov and J.~L. Lebowitz, J. Stat. Phys. {\bf 86},  953  (1997).

\bibitem{GaCo95a}
G. Gallavotti and E.~G.~D. Cohen, Phys. Rev. Lett. {\bf 74},  2694  (1995).

\bibitem{Ruelle96}
D. Ruelle, J. Stat. Phys. {\bf 85},  1  (1996).

\bibitem{ECM}
D.~J. Evans, E.~G.~D. Cohen, and G.~P. Morris, Phys. Rev. A {\bf 42},  5990
  (1990).

\bibitem{Vanc}
W.~N. Vance, Phys. Rev. Lett. {\bf 69},  1356  (1992).

\bibitem{BarEC}
A. Baranyai, D.~J. Evans, and E.~G.~D. Cohen, J. Stat. Phys. {\bf 70},  1085
  (1993).

\bibitem{DeGP95}
C. Dellago, L. Glatz, and H.~A. Posch, Phys. Rev. E {\bf 52},  4817  (1995).

\bibitem{Do99}
J.~R. Dorfman, {\em An introduction to chaos in nonequilibrium statistical
  mechanics} (Cambridge University Press, Cambrigde, 1999).

\bibitem{TGN98}
  {\em Chaos and Irreversibility}, Vol.~8 of {\em Chaos}, edited by T. T\'el,
  P. Gaspard, and G. Nicolis (American Institute of Physics, College Park,
  1998).

\bibitem{MaHo92}
  {\em Microscopic simulations of complex hydrodynamic phenomena}, Vol.~292
  of {\em NATO ASI Series B: Physics}, edited by M. Mareschal and B.~L. Holian
  (Plenum Press, New York, 1992).

\bibitem{Mare97}
  {\em The microscopic approach to complexity in non-equilibrium molecular
  simulations}, Vol.~240 of {\em Physica A}, edited by M. Mareschal (Elsevier,
  Amsterdam, 1997).

\bibitem{Ho88}
W.~G. Hoover {\it et~al.}, Phys. Lett. A {\bf 133},  114  (1988).

\bibitem{DeMo96}
C.~P. Dettmann and G.~P. Morriss, Phys. Rev. E {\bf 54},  2495  (1996).

\bibitem{DettmannMorriss97}
C.~P. Dettmann and G.~P. Morriss, Phys. Rev. E {\bf 55},  3693  (1997).

\bibitem{Choq98}
P. Choquard, Chaos {\bf 8},  350  (1998).

\bibitem{DePo97b}
C. Dellago and H.~A. Posch, J. Stat. Phys. {\bf 88},  825  (1997).

\bibitem{GN}
P. Gaspard and G. Nicolis, Phys. Rev. Lett. {\bf 65},  1693  (1990).

\bibitem{GaDo95}
P. Gaspard and J.~R. Dorfman, Phys. Rev. E {\bf 52},  3525  (1995).

\bibitem{DoGa95}
J.~R. Dorfman and P. Gaspard, Phys. Rev. E {\bf 51},  28  (1995).

\bibitem{Gasp}
P. Gaspard, {\em Chaos, Scattering, and Statistical Mechanics} (Cambridge
  University Press, Cambridge, 1998).

\bibitem{Lo05}
H.~A. Lorentz, Proc. Amst. Acad.  438  (1905).

\bibitem{BuSi81}
L.~A. Bunimovich and Y.~G. Sinai, Commun. Math. Phys. {\bf 78},  479  (1981).

\bibitem{CvGS92}
P. Cvitanovic, P. Gaspard, and T. Schreiber, Chaos {\bf 2},  85  (1992).

\bibitem{MaZw83}
J. Machta and R. Zwanzig, Phys. Rev. Lett. {\bf 50},  1959  (1983).

\bibitem{Gas93}
P. Gaspard, Chaos {\bf 3},  427  (1993).

\bibitem{GaBa94}
P. Gaspard and F. Baras, Phys. Rev. E {\bf 51},  5333  (1994).

\bibitem{MoRo94}
G.~P. Morriss and L. Rondoni, J. Stat. Phys. {\bf 75},  553  (1994).

\bibitem{Gasp96}
P. Gaspard, Phys. Rev. E {\bf 53},  4379  (1996).

\bibitem{MaMa97}
H. Matsuoka and R. Martin, J. Stat. Phys {\bf 88},  81  (1997).

\bibitem{LRM94}
J. Lloyd, L. Rondoni, and G.~P. Morriss, Phys. Rev. E {\bf 50},  3416  (1994).

\bibitem{LNRM95}
J. Lloyd, M. Niemeyer, L. Rondoni, and G.~P. Morriss, Chaos {\bf 5},  536
  (1995).

\bibitem{DettmannMorriss96}
C.~P. Dettmann and G.~P. Morriss, Phys. Rev. E {\bf 54},  4782  (1996).

\bibitem{DeMo97}
C.~P. Dettmann and G.~P. Morriss, Phys. Rev. Lett. {\bf 78},  4201  (1997).

\bibitem{RKrd}
K. Rateitschak and R. Klages (unpublished).

\bibitem{Ma1879}
J.~C. Maxwell, Cam. Phil. Trans. {\bf 12},  547  (1879).

\bibitem{Bo09}
L. Boltzmann,  in {\em Wissenschaftliche Abhandlungen von L. Boltzmann}, edited
  by F. Hasen\"ohrl (J. A. Barth Verlag, Leipzig, 1909), Vol.~2, Chap.~63.

\bibitem{MPT98}
L.~J. Milanovi\'c, H.~A. Posch, and W. Thirring, Phys. Rev. E {\bf 57},  2763
  (1998).

\bibitem{WKN98}
C. Wagner, R. Klages, and G. Nicolis, Phys. Rev. E {\bf 60}, (in press) (1999).

\bibitem{vK71}
N. van Kampen, Physica Norvegica {\bf 5},  279  (1971).

\bibitem{GaKl}
P. Gaspard and R. Klages, Chaos {\bf 8},  409  (1998).

\bibitem{BTV98}
W. Breymann, T. Tel, and J. Vollmer, Chaos {\bf 8},  396  (1998).

\bibitem{Groe95}
J. Groeneveld, priv.\ commun.

\bibitem{mapg}
R. Klages and J. Groeneveld, Verhandl. DPG {\bf VI},  646  (1998).

\bibitem{mapb}
R. Klages, Verhandl. DPG {\bf VI},  678  (1999).

\bibitem{RKj}
R. Klages (unpublished).

\bibitem{PoschHoover98}
H.~A. Posch and W.~G. Hoover, Phys. Rev. E {\bf 58},  4344  (1998).

\bibitem{RKHN99}
K. Rateitschak, R. Klages, W. Hoover, and G. Nicolis (unpublished).

\bibitem{LeSp98}
J. Lebowitz and H. Spohn, preprint cond-mat/9811220.

\bibitem{DePo95}
C. Dellago and H.~A. Posch, Phys. Rev. E {\bf 52},  2401  (1995).

\bibitem{DePH96}
C. Dellago, H.~A. Posch, and W.~G. Hoover, Phys. Rev. E {\bf 53},  1485
  (1996).

\bibitem{BSHP96}
N. Brillantov, F. Spahn, J.-M. Hertzsch, and T. P\"oschel, Phys. Rev. E {\bf
  53},  5382  (1996).

\bibitem{PoSch97}
T. P\"oschel and T. Schwager, Phys. Rev. Lett. {\bf 80},  5708  (1997).

\bibitem{SchPo98}
T. Schwager and T. P\"oschel, Phys. Rev. E {\bf 57},  650  (1998).

\bibitem{GiZi96}
G. Giese and A. Zippelius, Phys. Rev. E {\bf 54},  4828  (1996).

\bibitem{AGZ98}
T. Aspelmeier, G. Giese, and A. Zippelius, Phys. Rev. E {\bf 57},  857  (1998).

\end{thebibliography}
\end{document}